\documentclass[aps,pra,twocolumn,showpacs]{revtex4}

\usepackage[dvips]{graphicx}

\input{epsf}

\begin{document}

\title{Momentum distribution and condensate fraction of a Fermi gas in the BCS-BEC crossover}

\author{G. E. Astrakharchik$^{(1,2)}$, J. Boronat$^{(3)}$, J. Casulleras$^{(3)}$, and S. Giorgini$^{(4,1)}$}
\address{$^{(1)}$Dipartimento di Fisica, Universit\`a di Trento and BEC-INFM, I-38050 Povo, Italy\\
$^{(2)}$Institute of Spectroscopy, 142190 Troitsk, Moscow region, Russia\\
$^{(3)}$Departament de F\'{\i}sica i Enginyeria Nuclear, Campus Nord B4-B5, Universitat Polit\`ecnica de Catalunya,
E-08034 Barcelona, Spain\\
$^{(4)}$ JILA, University of Colorado, Boulder, CO 80309-0440, U.S.A.}

\date{\today}

\begin{abstract}
By using the diffusion Monte Carlo method  we calculate the one- and two-body 
density matrix of an interacting Fermi gas at $T=0$ in the BCS-BEC crossover. 
Results for the momentum distribution of the atoms, as obtained from  the 
Fourier transform of the one-body density matrix, are reported as a function 
of the interaction strength. Off-diagonal long-range order in the system is 
investigated through the asymptotic behavior of the two-body density matrix. 
The condensate fraction of fermionic pairs is calculated in the unitary limit 
and on both sides of the BCS-BEC crossover.  
\end{abstract}

\pacs{}

\maketitle
The physics of the crossover from Bardeen-Cooper-Schrieffer (BCS)
superfluids to molecular Bose-Einstein condensates (BEC)  in ultracold
Fermi gases near a Feshbach resonance is a very exciting field that has
recently attracted a lot of interest, both from the
experimental~\cite{EXP1,EXP2} and the theoretical side~\cite{Levin}. An
important experimental achievement is the observation of a condensate of
pairs of fermionic atoms on the side of the Feshbach resonance where no
stable  molecules would exist in vacuum~\cite{EXP3,EXP4}. Although the
interpretation of the experiment is not straightforward, as  it involves
an out-of-equilibrium projection technique of fermionic pairs onto bound
molecules~\cite{TH}, it is believed  that these results strongly support
the existence of a superfluid order parameter in the strongly correlated
regime on the  BCS side of the resonance~\cite{EXP4}. 

The occurrence of off-diagonal long-range order (ODLRO) in interacting
systems of bosons and fermions was investigated by  C.N. Yang in terms of
the asymptotic behavior of the one- and two-body density
matrix~\cite{Yang}. In the case of a  two-component Fermi gas with
$N_\uparrow$ spin-up and $N_\downarrow$ spin-down particles, the one-body
density  matrix (OBDM) for spin-up particles, defined as
\begin{equation}
\rho_1({\bf r}_1^\prime,{\bf r}_1)=\langle\psi_\uparrow^\dagger({\bf r}_1^\prime)
\psi_\uparrow({\bf r}_1)\rangle \;,
\label{OBDM}
\end{equation}     
does not possess any eigenvalue of order $N_\uparrow$. This behavior 
implies for homogeneous systems the asymptotic condition 
$\rho_1({\bf r}_1^\prime,{\bf r}_1)\to 0$ as $|{\bf r}_1-{\bf
r}_1^\prime|\to\infty$.
In the above expression
$\psi_\uparrow^\dagger({\bf r})$ ($\psi_\uparrow({\bf r})$) denote the
creation (annihilation)  operator of spin-up particles. The same result
holds for spin-down particles. ODLRO may occur instead in the two-body 
density matrix (TBDM), that is defined as
\begin{equation}
\rho_2({\bf r}_1^\prime,{\bf r}_2^\prime,{\bf r}_1,{\bf r}_2)=
\langle\psi_\uparrow^\dagger({\bf r}_1^\prime)
\psi_\downarrow^\dagger({\bf r}_2^\prime)\psi_\uparrow({\bf r}_1)
\psi_\downarrow({\bf r}_2)\rangle \;.
\label{TBDM}
\end{equation}      
For an unpolarized gas with $N_\uparrow=N_\downarrow=N/2$, 
if $\rho_2$ has an eigenvalue of the order of the total number of
particles $N$, the TBDM can be  written as a
spectral decomposition separating the largest eigenvalue,
\begin{equation}
\rho_2({\bf r}_1^\prime,{\bf r}_2^\prime,{\bf r}_1,{\bf r}_2)=
\alpha N/2 \varphi^\ast({\bf r}_1^\prime,{\bf r}_2^\prime)
\varphi({\bf r}_1,{\bf r}_2) + \rho_2^\prime \;,
\label{TBDM1}
\end{equation}
$\rho_2^\prime$ containing only eigenvalues of order one.
The parameter $\alpha\le 1$ in Eq.
(\ref{TBDM1}) is interpreted as the condensate  fraction of pairs, in a
similar way as the condensate fraction of single atoms is derived from the OBDM. 

The
spectral decomposition (\ref{TBDM1}) yields for homogeneous systems the
following asymptotic  behavior of the TBDM
\begin{equation}
\rho_2({\bf r}_1^\prime,{\bf r}_2^\prime,{\bf r}_1,{\bf r}_2) \to \alpha N/2 
\varphi^\ast(|{\bf r}_1^\prime-{\bf r}_2^\prime|)
\varphi(|{\bf r}_1-{\bf r}_2|) \;,
\label{TBDM2}
\end{equation}        
if $|{\bf r}_1-{\bf r}_1^\prime|$, $|{\bf r}_2-{\bf
r}_2^\prime|\to\infty$. The complex function $\varphi$ is 
proportional to the order parameter $\langle\psi_\uparrow({\bf
r}_1)\psi_\downarrow({\bf r}_2)\rangle=\sqrt{\alpha N/2} \varphi(|{\bf
r}_1-{\bf r}_2|)$, whose appearance distinguishes the superfluid state of the
Fermi gas.
Equation (\ref{TBDM2}) should be
contrasted with the behavior of Bose systems with ODLRO, where $\rho_1$
has an eigenvalue  of order $N$~\cite{Penrose}, and consequently the
largest eigenvalue of $\rho_2$ is of the order of $N^2$.

In this Letter we present fixed-node diffusion Monte Carlo (FN-DMC)
results of $\rho_1$ and $\rho_2$ for a homogeneous interacting 
Fermi  gas at $T=0$ in
the BCS-BEC crossover. From the Fourier transform of $\rho_1(r)$, we
calculate the momentum distribution of the  gas, $n_{\bf k} = \int d^3{\bf r}
\rho_1(r) e^{i{\bf k}\cdot{\bf r}}$, as a function of the interaction
strength. From the asymptotic behavior of $\rho_2$, we extract the value of the 
condensate fraction of pairs $\alpha$.
The calculated condensate 
fraction is compared with analytical expansions holding on the BEC and BCS
side of the Feshbach resonance. The comparison with  mean-field
results~\cite{BCS-BEC} for $n_{\bf k}$ and $\alpha$ in the crossover region is
also discussed.

We consider a homogeneous two-component unpolarized Fermi gas described by the Hamiltonian
\begin{equation}
H=-\frac{\hbar^2}{2m}\left( \sum_{i=1}^{N/2}\nabla^2_i + 
\sum_{i^\prime=1}^{N/2}\nabla^2_{i^\prime}\right)
+\sum_{i,i^\prime}V(r_{ii^\prime}) \;,
\label{hamiltonian}
\end{equation}   
where $m$ is the mass of the particles and $i,j,...$
($i^\prime,j^\prime,...$) label spin-up (spin-down) particles.  
The strength of the interaction is assumed to be determined only by
the parameter $1/k_Fa$, with $k_F=(3\pi^2n)^{1/3}$  the  Fermi wave-vector
fixed by the atomic density $n=N/V$, and $a$ the s-wave scattering length 
describing the low-energy  collisions between the two fermionic species.
The interatomic interactions in Eq. (\ref{hamiltonian}) are only
between atoms with different spin and are modeled by a 
short-range potential that determines the value of $a$. In the present
study, we use an attractive square-well potential, $V(r)=-V_0$ for $r<R_0$
and $V(r)=0$ otherwise, with $nR_0^3=10^{-6}$. We have verified that
in the density range
$nR_0^3=10^{-7}- 10^{-5}$ the particular form of $V(r)$ 
is not relevant, and therefore the present
results are in this sense universal. 
The different regimes: BEC
($a>0$ and $1/k_Fa\gg 1$), BCS ($a<0$ and $1/k_F|a|\gg 1$) and unitary
limit ($1/k_Fa=0$), are obtained by varying the potential depth $V_0$ as 
in~\cite{US}. Quantum Monte Carlo studies of the Hamiltonian
(\ref{hamiltonian}) have already been carried out to investigate the $T=0$
equation of state~\cite{QMC1,US} and the pairing gap~\cite{QMC1}. 

In a FN-DMC simulation, the wave function $f({\bf R},\tau)=\psi_T({\bf
R})\Psi({\bf R},\tau)$  (${\bf R}={\bf r}_1,\ldots,{\bf
r}_{N_\uparrow},{\bf r}_{1^\prime},\ldots,{\bf r}_{N_\downarrow}$) is evolved 
in imaginary  time $\tau=it/\hbar$ according to the time-dependent
Schr\"odinger equation, with $\psi_T({\bf R})$ acting as importance 
sampling function and as nodal constraint. The function $\Psi({\bf R})
\equiv\Psi({\bf R},\tau\to\infty)$, which is the lowest energy state of the system 
having the nodes of $\psi_T({\bf R})$, is obtained from the large-time
evolution of $f({\bf R},\tau)$.
The trial wave function we consider has the general form~\cite{QMC1,QMC2}
$\psi_T({\bf R})=\Psi_{\text{J}}({\bf R}) \Psi_{\text{BCS}}({\bf R})$, where
$\Psi_{{\text J}}$ contains Jastrow correlations between all the particles, 
\begin{equation}
\Psi_{{\text J}}({\bf R})=\prod_{i<j}f_{\uparrow\uparrow}(r_{ij})
\prod_{i^\prime<j^\prime}f_{\downarrow\downarrow}
(r_{i^\prime j^\prime})\prod_{i,i^\prime}f_{\uparrow\downarrow}(r_{ii^\prime}) \;,
\label{trialwf1}
\end{equation}  
and the BCS-type wave function $\Psi_{{\text BCS}}$ is the antisymmetrized product
of the pair wave functions  $\phi({\bf r}_i-{\bf r}_{i^\prime})$ 
\begin{equation}
\Psi_{\text{BCS}}({\bf R})={\cal A} \left( \phi({\bf r}_1-{\bf r}_{1^\prime})
\phi({\bf r}_2-{\bf r}_{2^\prime})...
\phi({\bf r}_{N_\uparrow}-{\bf r}_{N_\downarrow})\right) \;.
\label{trialwf2}
\end{equation}
The pair orbital $\phi({\bf r})$ is chosen as
\begin{equation}
\phi({\bf r})=\beta\sum_{k_\alpha\le k_{max}}e^{i{\bf k}_\alpha \cdot {\bf r}} + 
\phi_s(r) \;,
\label{trialwf3}
\end{equation}
where the sum is performed over the plane wave states ${\bf
k}_\alpha=2\pi/L(\ell_{\alpha x}\hat{x}+\ell_{\alpha y}\hat{y}
+\ell_{\alpha z}\hat{z})$ up to the largest closed shell
$k_{max}=2\pi/L(\ell_{max x}^2+\ell_{max y}^2+\ell_{max z}^2)^{1/2}$
occupied by $N/2$ particles. 
Here $L=V^{1/3}$ is the size of the cubic simulation box and $\ell$ are
integer numbers. If $\phi_s(r)=0$ in Eq. (\ref{trialwf3}), $\Psi_{\text{BCS}}$ 
in Eq. (\ref{trialwf2}) coincides with the exact wave function of a free
Fermi gas, i.e., the product of Slater
determinants of spin-up and spin-down particles~\cite{Bouchaud}.
The spherically symmetric function $\phi_s(r)$ in Eq. (\ref{trialwf3})
accounts for $s$-wave pairing. If  $1/k_Fa\ge -0.2$, $\phi_s({\bf r})$
corresponds to the solution of the two-body problem with the potential $V(r)$, 
as in Ref.~\cite{US}.
For $1/k_Fa<-0.2$ we use instead $\phi_s(r)=
\gamma_1[\exp(-\gamma_2 r)+\exp(-\gamma_2(L-r))]$.
Here, $\gamma_1$, $\gamma_2$, and $\beta$ in Eq. (\ref{trialwf3}) are variational
parameters. 

The Jastrow  wave function
$\Psi_{{\text J}}$, Eq. (\ref{trialwf1}), is determined as follows.  For
$1/k_Fa\ge -0.2$, we use  $f_{\uparrow\downarrow}(r)=1$ and
$f_{\uparrow\uparrow}(r)=f_{\downarrow\downarrow}(r)$ 
given by the two-body solution of a fictitious
repulsive step potential  
with range $\tilde{R}$ and scattering length $\tilde{a}$.
The boundary conditions $f(r=L/2)=1$ and $f^\prime(r=L/2)=0$ determine the wave function
in terms of the variational parameters $\tilde{R}$ and $\tilde{a}$.   
For  $1/k_Fa< -0.2$, we use
instead $f_{\uparrow\uparrow}(r)= f_{\downarrow\downarrow}(r) =1$ and the
model used in Ref.~\cite{US}  for the crossed correlation factor,
$f_{\uparrow\downarrow}(r)$. 
It is worth noticing that the function 
$\psi_T({\bf R})$ defined above reproduces as a special case the trial
wave function used in the preceding study~\cite{US}, but contains more
variational parameters. The parameters of the Jastrow function $\Psi_{{\text J}}$, 
Eq. (\ref{trialwf1}), are optimized by minimizing the variational expectation value
$\langle\psi_T|H|\psi_T\rangle/\langle\psi_T|\psi_T\rangle$. The parameters of the BCS 
function $\Psi_{\text{BCS}}$, Eq. (\ref{trialwf2}), affect the nodal surface of the trial 
wave function and they are optimized by minimizing the FN-DMC estimate of the energy.
For the values of $1/k_Fa$ used in the present study, the calculated FN-DMC energies 
are in agreement with the results obtained in \cite{US}, although the optimized 
variational energy has significantly improved.

A direct estimate of any operator $O$ in DMC  is known as mixed
estimate, $\langle
O\rangle_{\text{m}}=\langle\psi_T|O|\Psi\rangle/\langle\psi_T|\Psi\rangle$,
and is exact only for the Hamiltonian and operators commuting with it. If
$O$ is a local operator, one can circumvent this problem by introducing
pure estimators. That is not the case for $\rho_1$ and $\rho_2$ which are
the objectives of the present work. In order to reduce, and even eliminate
in practice, a possible bias in the calculation we have used the
extrapolated estimator $\langle\Psi|O|\Psi\rangle/\langle\Psi|\Psi\rangle
\simeq 2\langle
O\rangle_{\text{m}}-\langle O\rangle_{\text{v}}$, with $\langle
O\rangle_{\text{v}}=\langle\psi_T|O|\psi_T\rangle/\langle\psi_T|\psi_T\rangle$.
Any residual bias in the extrapolated estimator is reduced if the trial  
function $\psi_T({\bf R})$  has a large overlap with $\Psi({\bf R})$.
Consequently, compared to the calculation of
the  eigenenergies, the optimization of $\psi_T$ is a more important
issue in the calculation of observables like the OBDM  and
TBDM~\cite{QMC2}.

\begin{figure}
\begin{center}
\includegraphics*[width=7.5cm]{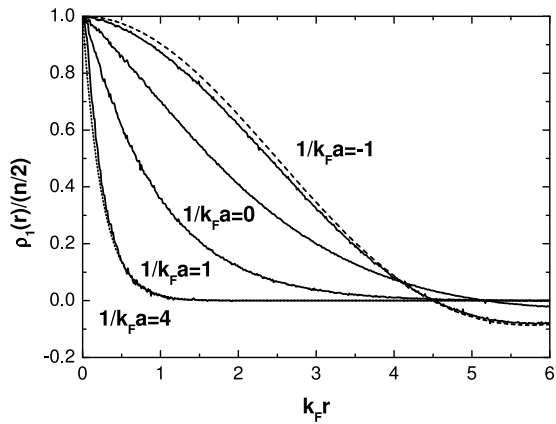}
\caption{OBDM for different values of the interaction strentgth $1/k_Fa$ (solid lines). 
The dotted line corresponds to 
$e^{-r/a}$ for $1/k_Fa=4$ and the dashed line is the OBDM of a non interacting gas.}
\label{fig1}
\end{center}
\end{figure}

\begin{figure}
\begin{center}
\includegraphics*[width=7.5cm]{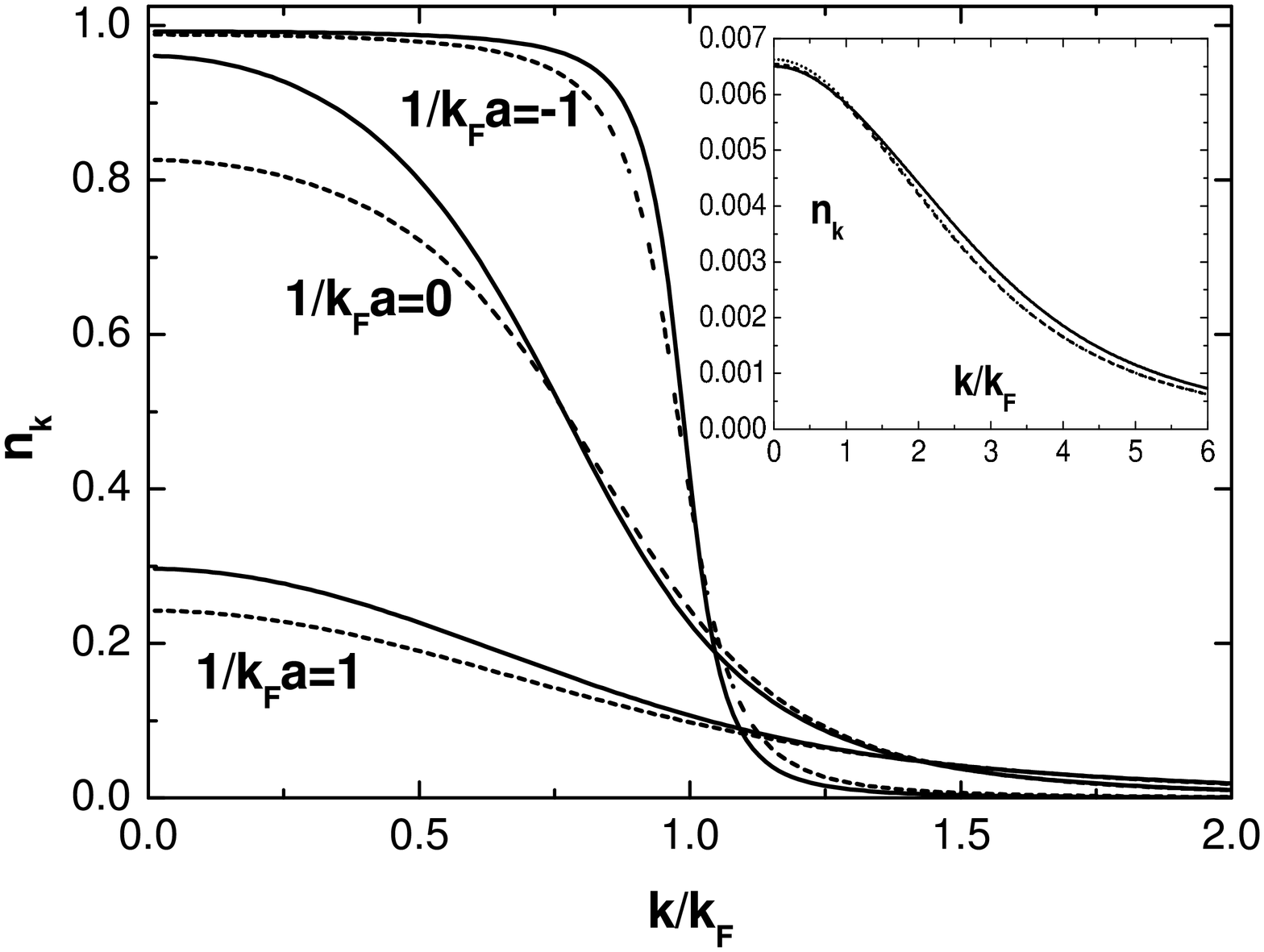}
\caption{Momentum distribution $n_{\bf k}$ for different values of $1/k_Fa$ (solid lines). 
The dashed lines correspond to $n_{\bf k}$ 
calculated using the BCS mean-field approach~\cite{Viverit}. 
Inset: $n_{\bf k}$ for $1/k_Fa=4$. The dotted line corresponds to the momentum distribution 
of the molecular state (see text). It almost coincides with the mean-field result (dashed line).}
\label{fig2}
\end{center}
\end{figure}

We consider a system with $N=66$ particles and periodic boundary
conditions. In Fig.~\ref{fig1} we show results of the OBDM $\rho_1(r)$,
Eq. (\ref{OBDM}), for different values  of the interaction strength
$1/k_Fa$. In the deep molecular regime,  $1/k_Fa\gg 1$, the OBDM is
determined by the molecular wave function $\phi_{\text{bs}}(r)$: $\rho_1(r)\simeq
n/2\int  d^3{\bf r}^\prime \phi_{\text{bs}}^\ast(|{\bf r}+{\bf
r}^\prime|)\phi_{\text{bs}}(r^\prime)$. For a zero-range potential the molecular
wave  function is given by $\phi_{\text{bs}}(r)\propto e^{-r/a}/r$ and one finds
$\rho_1(r)\simeq n e^{-r/a}/2$. This behavior is shown in  Fig.~\ref{fig1}
for $1/k_Fa=4$. If one moves closer to the resonance, the
OBDM decays slowly and oscillations start to appear. Finally, on the BCS
side of the resonance, the OBDM becomes more and more similar to the ideal
gas result $\rho_1(r)= 3n[\sin(k_Fr)/(k_Fr)-\cos(k_Fr)]/[2(k_Fr)^2]$. The
momentum distribution $n_{\bf k}$, obtained from the Fourier transform  of
$\rho_1(r)$, is shown in Fig.~\ref{fig2}. In the inset of Fig.~\ref{fig2}
we compare $n_{\bf k}$, calculated using FN-DMC for  $1/k_Fa=4$, with the
momentum distribution of the atoms in the molecular state
$n_{\bf k}=4(k_Fa)^3/[3\pi(1+k^2a^2)^2]$~\cite{Viverit}.  To reduce finite-size
effects in the calculation of the Fourier transform for $1/k_Fa=0$ and
$-1$, we have used the  following model for the $k$-dependence of $n_{\bf k}$
\begin{equation}
n_{\bf k}=A\left(1-\frac{(k/k_F)^2-\mu}{\sqrt{[(k/k_F)^2-\mu]^2+\Delta^2}} \right) \;.
\label{modelnk}
\end{equation}
The values of $\mu$, $\Delta$ and $A$ are free  parameters determined by
the best fit of the inverse Fourier transform of Eq. (\ref{modelnk}) to the
calculated $\rho_1(r)$, with the  constraint $1/V\sum_k n_{\bf k}=n/2$. 
For $A=1/2$, the above expression reproduces the standard $n_{\bf k}$ of BCS theory with 
$\mu$  and $\Delta$, respectively, chemical potential and gap in units of the Fermi
energy $\epsilon_F=\hbar^2k_F^2/2m$. For $1/k_Fa=0$ and -1 the Fourier transform of the BCS 
model, Eq. (\ref{modelnk}), reproduces quite well the calculated $\rho_1(r)$ with a $\chi^2/\nu$
of the order of one. In Fig.~\ref{fig2} we also show the results of $n_{\bf k}$ obtained using 
the BCS mean-field  theory~\cite{BCS-BEC}, where the values of chemical potential
and gap are calculated self-consistently through the gap and  number
equations~\cite{Viverit}. As evident from Fig.~\ref{fig2}, the mean-field theory sistematically 
overestimates the broadening of the momentum distribution, apart from the deep BEC regime where
both mean-field and FN-DMC results coincide with the momentum distribution of the molecular state.

\begin{figure}
\begin{center}
\includegraphics*[width=7.5cm]{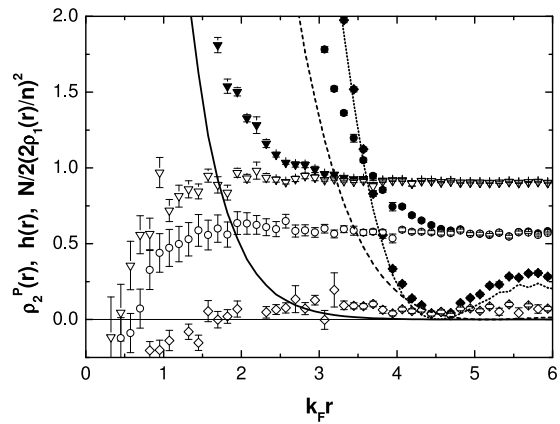}
\caption{Projected TBDM, $\rho_2^P(r)$, (solid symbols), $N/2(2\rho_1(r)/n)^2$ (lines) 
and $h(r)=\rho_2^P(r)-N/2(2\rho_1(r)/n)^2$
(open symbols) for different values of $1/k_Fa$: $1/k_Fa=1$ 
(solid line and triangles), $1/k_Fa=0$ (dashed line and circles) 
and $1/k_Fa=-1$ (dotted line and diamonds).}
\label{fig3}
\end{center}
\end{figure}

\begin{figure}
\begin{center}
\includegraphics*[width=7.5cm]{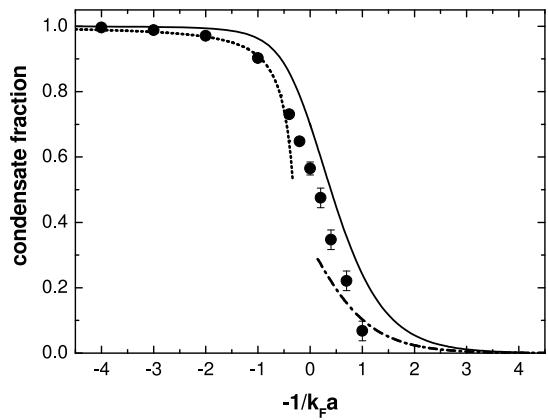}
\caption{Condensate fraction of pairs $\alpha$ as a function of the interaction 
strength: FN-DMC results (symbols), Bogoliubov 
quantum depletion of a Bose gas with $a_m=0.6 a$ (dashed line), BCS theory 
using Eq.~(\ref{FBCS}) (dot-dashed line) and 
self-consistent mean-field theory (solid line).}
\label{fig4}
\end{center}
\end{figure}

The condensate fraction of pairs has been obtained from the 
the projected TBDM, defined as~\cite{Senatore,Ortiz} 
\begin{equation}
\rho_2^P(r)=\frac{2}{N}\int d^3{\bf r}_1 d^3{\bf r}_2 
\rho_2({\bf r}_1+{\bf r},{\bf r}_2+{\bf r},{\bf r}_1,{\bf r}_2) \;.
\label{PTBDM}
\end{equation}      
Assuming the eigenvector $\varphi(r)$ of Eq. (\ref{TBDM2}) normalized to
$1/V$, the large $r$ behavior of $\rho_2^P(r)$ tends to the condensate
fraction $\alpha$: $\lim_{r\to\infty}\rho_2^P(r)=\alpha$. In terms of the
order parameter, $F(|{\bf r}_1-{\bf r}_2|) =\langle\psi_\uparrow({\bf
r}_1)\psi_\downarrow({\bf r}_2)\rangle$, one has instead
$\lim_{r\to\infty}\rho_2^P(r)=2/n\int  d^3{\bf r}^\prime |F(r^\prime)|^2$.
In Fig.~\ref{fig3} we show the results of $\rho_2^P(r)$. Finite size
effects can be  substantially reduced if one considers the following
decomposition: $\lim_{r\to\infty} \rho_2({\bf r}_1+{\bf r},{\bf r}_2+{\bf
r},{\bf r}_1,{\bf r}_2)=|F(|{\bf r}_1-{\bf r}_2|)|^2+\rho_1^2(r)$,
accounting for the large  $r$ behavior of $\rho_2$ when $\rho_1(r)$ is
small  but not zero. From this result one finds for the asymptotic behavior
of the projected TBDM: $\lim_{r\to\infty}\rho_2^P(r)=\alpha  +
N/2(2\rho_1(r)/n)^2$. In Fig.~\ref{fig3} we show results for the quantity
$h(r)=\rho_2^P(r)-N/2(2\rho_1(r)/n)^2$. For values  of $1/k_Fa$ on the BEC
side of the Feshbach resonance and up to the unitary limit $1/k_Fa=0$, the
asymptotic values of $\rho_2(r)$  and $h(r)$ coincide. On the BCS side,
instead, finite-size effects become visible in the large $r$ behavior of
$\rho_2^P(r)$, but are strongly suppressed for $h(r)$. The results for the
condensate fraction $\alpha$, as obtained from the asymptotic behavior of
$h(r)$, are shown in Fig.~\ref{fig4}. In the BEC regime, the results
reproduce the Bogoliubov quantum depletion of a gas of  composite bosons
$\alpha=1-8\sqrt{n_m a_m^3}/3\sqrt{\pi}$, where $n_m=n/2$ is the density of
molecules and $a_m=0.6 a$ is the  dimer-dimer scattering
length~\cite{Petrov,US}. In the opposite BCS regime, the condensate
fraction $\alpha$ can be calculated from the result of the BCS order
parameter holding for $r\gg a$~\cite{BCS}
\begin{equation}
F_{\text{BCS}}(r)=\frac{\Delta k_F^3}{\epsilon_F}\frac{\sin(k_Fr)}{4\pi^2 k_Fr}K_0(r/\xi_0) \;,
\label{FBCS}
\end{equation}     
where $\xi_0=\hbar^2k_F/m\Delta$ is the coherence length and $K_0(x)$ is
the modified Bessel function. If we include the Gorkov-Melik Barkhudarov 
correction for the
pairing gap~\cite{Gorkov} $\Delta=(2/e)^{7/3}\epsilon_F e^{-\pi/2k_F|a|}$,
we obtain for $\alpha=2/n\int  d^3{\bf r}F_{\text{BCS}}^2(r)$ the dot-dashed line
shown in Fig.~\ref{fig4}. For $1/k_Fa\le -1$ the coherence length $\xi_0$ becomes increasingly 
larger and finite-size effects start to be relevant in the FN-DMC estimate of $\alpha$. For example 
at $1/k_Fa=-1$, the value of $\alpha$, as obtained from $F_{BCS}(r)$ [Eq. (\ref{FBCS})], reduces 
from 0.10 to 0.08 by cutting off the spacial integral at the simulation box size $r=L/2$.   
The condensate fraction $\alpha$ can  also be
estimated in the crossover region using the self-consistent mean-field
approach~\cite{BCS-BEC} and the result for the  order parameter:
$F(r)=1/V\sum_{\bf k}u_k v_k e^{i{\bf k}\cdot{\bf r}}$, where $u_k$ and
$v_k$ are the usual quasiparticle  amplitudes of BCS theory. The result is
shown in Fig.~\ref{fig4} with a solid line (see also Ref.~\cite{Ortiz}).
The mean-field  result reproduces the qualitative behavior of the
condensate fraction across the resonance. However, it does not reproduce
the quantum depletion in the BEC regime, nor the Gorkov-Melik 
Barkhudarov correction to the gap in the BCS regime.
            
In conclusion, we have investigated using quantum Monte Carlo techniques
the one- and two-body density matrix of a  homogeneous Fermi gas
in the BCS-BEC crossover. Results for the momentum distribution are
obtained as a function of the  interaction strength. The condensate
fraction of pairs is calculated from the asymptotic behavior of the
two-body density matrix. The results obtained are in good agreement with
the quantum depletion of a weakly-interacting Bose gas in the BEC  regime
and with the normalization of the BCS order parameter in the BCS regime.

Acknowledgements:GEA and SG acknowledge 
support by the Ministero dell'Istruzione, dell'Universit\`a e della Ricerca (MIUR). 
JB and JC acknowledge support from DGI 
(Spain) Grant No. BFM2002-00466 and Generalitat de Catalunya Grant No. 2001SGR-00222.

\end{document}